\documentstyle[twocolumn,aps,epsfig,floats]{revtex}

\begin{document}

\twocolumn[\hsize\textwidth\columnwidth\hsize\csname %
@twocolumnfalse\endcsname

\title{The Resonance Peak in Sr$_2$RuO$_4$:
Signature of Spin Triplet Pairing}
\author{Dirk K. Morr$^1$, Peter F. Trautman$^{1,2}$, and Matthias J. Graf$^1$ }
\address{$^1$ Theoretical Division, MS B262, Los Alamos National
Laboratory, Los Alamos, NM 87545\\
$^2$ Baylor University, Waco, TX 76798}
\date{\today}
\draft
\maketitle

\begin{abstract}
We study the dynamical spin susceptibility,
$\chi({\bf q}, \omega)$, in the normal and superconducting state
of Sr$_2$RuO$_4$. In the normal state, we find a peak in the
vicinity of ${\bf Q}_i\simeq (0.72\pi,0.72\pi)$ in agreement with
recent inelastic neutron scattering (INS) experiments. We predict
that for spin triplet pairing in the superconducting state a {\it
resonance peak} appears in the out-of-plane
component of $\chi$, but is absent in the in-plane component. In contrast,
no resonance peak is expected for spin singlet pairing.

\end{abstract}

\pacs{PACS numbers: 74.25.-q, 74.25.Ha, 74.70.-b}

]

The superconducting (SC) state of Sr$_2$RuO$_4$ has been the focus
of intense experimental and theoretical research over the last few
years. Sr$_2$RuO$_4$ is isostructural with the high-temperature
superconductor (HTSC) La$_{2-x}$Sr$_x$CuO$_4$ and is the only
known layered perovskite which is superconducting in the absence
of Cu \cite{maeno94}. Understanding the pairing mechanism in
Sr$_2$RuO$_4$ could therefore provide important insight into the
origin of unconventional superconductivity in general, and that of
the HTSC in particular. Since a related compound, SrRuO$_3$, is a
ferromagnet, it was suggested \cite{Rice95,Bas96} that
Sr$_2$RuO$_4$ is a triplet superconductor in which the pairing is
mediated by ferromagnetic paramagnons. Experimental support for
spin triplet pairing comes from Knight shift (KS) \cite{Muk99} and
elastic neutron scattering (ENS) measurements \cite{Duf00}, while
$\mu$SR \cite{Luke98} provides evidence for a broken time-reversal
symmetry in the SC state. However, the momentum dependence of the
superconducting gap is still unclear. While originally a $p$-wave
symmetry, belonging to the $E_u$ representation of the $D_{4h}$
point group, was proposed for the superconducting gap
\cite{Rice95,agterberg97}, $\Delta({\bf k}) \sim k_x + i k_y$,
recent specific heat \cite{Nis00}, thermal conductivity
\cite{thermal}, penetration depth \cite{bonalde00}, and nuclear
magnetic resonance \cite{Ish00} experiments suggest the presence
of line nodes in $\Delta({\bf k})$ and thus pairing with higher
orbital momentum.

The spin susceptibility, $\chi({\bf q}, \omega)$, is an important
input parameter for any theory ascribing the pairing
mechanism in Sr$_2$RuO$_4$ to the exchange of spin fluctuations.
In this letter we present a scenario for the momentum and frequency
dependence of $\chi({\bf q}, \omega)$,
both in the normal and superconducting
state. In the normal state, we find a peak in Im$\, \chi$ whose
momentum position is close to that reported by Sidis {\it et
al.}~\cite{Sid99} in inelastic neutron scattering (INS)
experiments. Our results for Re$\, \chi$ agree with the prediction
by Mazin and Singh \cite{Maz99} of a peak in the normal-state
static susceptibility, $\chi({\bf q}, \omega=0)$, around ${\bf
q}=(2\pi/3,2\pi/3)$.  We show that for triplet pairing in the
superconducting state the in-plane, $\chi_{\pm}=(\chi_{xx} +
\chi_{yy})/2$, and out-of-plane, $\chi_{zz}$, components of the
dynamic spin susceptibility are qualitatively different. In
particular, we predict that a {\it resonance peak}, similar to the
one observed in the HTSC \cite{exp_respeak}, appears in
$\chi_{zz}$, but is absent in $\chi_{\pm}$. Since no resonance
peak
exists for spin singlet pairing, it is an important signature of
spin triplet superconductivity.

\begin{figure} [t]
\begin{center}
\leavevmode
\epsfxsize=5cm
\epsffile{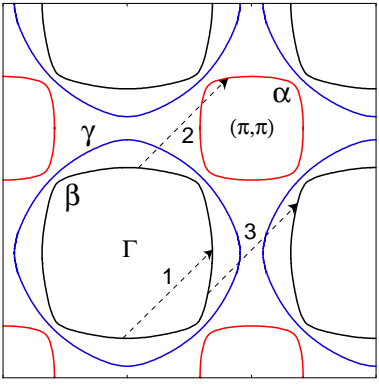}
\end{center}
\caption{Fermi surfaces of Sr$_2$RuO$_4$ in the extended Brillouin zone.
The arrows show quasiparticle excitations with nesting wavevector
${\bf Q}_i \simeq (0.72\pi, 0.72\pi)$ where we set the lattice constant $a=1$.}
\label{FS}
\end{figure}

Contributions to the dynamic spin susceptibility in Sr$_2$RuO$_4$
come from three electronic bands which are derived from the Ru 4d
$xy$, $xz$, and $yz$-orbitals. A comparison of angle-resolved
photoemission (ARPES) \cite{Puc98} and de Haas-van Alphen (dHvA)
\cite{Mac96} experiments with band-structure calculations
\cite{LDA} shows a substantial hybridization only between the
$xz$- and $yz$-orbitals, with a resulting hole-like
($\alpha$-band) and  electron-like Fermi surface (FS)
($\beta$-band), while the decoupled $xy$-orbitals give rise to the
electron-like $\gamma$-band \cite{com2}. Thus the electronic
structure of Sr$_2$RuO$_4$ can be described by the tight-binding
Hamiltonian
\begin{eqnarray}
{\cal H} &=&  \sum_{{\bf k}, \sigma} \epsilon^{xy}_{\bf k}
c^\dagger_{{\bf k},\sigma} c_{{\bf k},\sigma} + \sum_{{\bf k},
\sigma} \epsilon^{xz}_{\bf k} a^\dagger_{{\bf k},\sigma} a_{{\bf
k} ,\sigma} \nonumber \\
 & & \quad  + \sum_{{\bf k}, \sigma} \epsilon^{yz}_{\bf k}
b^\dagger_{{\bf k},\sigma} b_{{\bf k},\sigma} - \sum_{{\bf k},
\sigma} \left( t_\perp a^\dagger_{{\bf k},\sigma} b_{{\bf
k},\sigma} + h.c. \right)
\, ,
\label{elH}
\end{eqnarray}
where $c^\dagger_k, a^\dagger_k, b^\dagger_k$ are the fermionic
creation operators in the ${xy}, {xz}$, and ${yz}$-bands, with
spin $\sigma$,
respectively. The normal state tight-binding dispersions are given
by \cite{LDA}
\begin{eqnarray}
&
\epsilon^{i}_{\bf k} = -2t_x \cos k_x -2t_y \cos k_y
+4t^\prime \cos k_x \cos k_y  -\mu  \,,
&
\end{eqnarray}
with $(t_x, t_y, t^\prime, \mu)=$(0.44, 0.44, -0.14, 0.50)eV,
(0.31, 0.045, 0.01, 0.24)eV, (0.045, 0.31, 0.01, 0.24)eV for the
$i={xy}, {xz}$, ${yz}$-bands, respectively, and $t_\perp$
reflecting the hybridization between the ${xz}$ and ${yz}$-bands
\cite{com1}. After diagonalizing the Hamiltonian, Eq.(\ref{elH}),
we obtain the energy dispersions for the $\gamma$ and hybridized $\alpha$ and
$\beta$-bands
\begin{eqnarray}
\epsilon_{\alpha,\beta}({\bf k}) &=& \epsilon^{+}_{\bf k}
\mp \sqrt{ ({\epsilon^{-}_{\bf k}})^2+t_\perp^2 }
\quad , \
 \epsilon_{\gamma}({\bf k})=\epsilon^{xy}_{\bf k} \ ,
\label{abg_disp}
\end{eqnarray}
with $\epsilon^{\pm}_{\bf k} = (\epsilon^{xz}_{\bf k} \pm \epsilon^{yz}_{\bf
k})/2$. By fitting the area and
shape of the $\alpha$ and $\beta$-FS to those observed by ARPES
\cite{Puc98} and dHvA experiments \cite{Mac96}, we obtain $t_\perp
\approx 0.1$ eV; the Fermi surfaces for all three bands are shown in
Fig.~\ref{FS}.

%
%

The superconducting gap for unitary spin triplet pairing can be
written as
\begin{equation}
{\bf \Delta}_{\zeta \eta}({\bf k})=
 [{\bf d}({\bf k})\cdot \sigma
 \, i \, \sigma_2]_{\zeta \eta}
\label{gap}
\end{equation}
where $\sigma$ are the Pauli matrices. We assume that spin-orbit
coupling locks the {\bf d} vector along the crystal $\hat{c}$
axis, {\it i.e.}, ${\bf d} || \hat{z} || \hat{c}$, consistent with
KS \cite{Muk99} and ENS \cite{Duf00} experiments. In the
following, we consider a superconducting gap with `$f_{xy}$-wave'
($E_u$) symmetry,
\begin{equation}
d_z({\bf k})=\Delta({\bf k})=\Delta_0
\sin{k_x}\sin{k_y}\left(\sin{k_x}+i \sin{k_y}\right) \ ,
\label{fwave}
\end{equation}
which was shown \cite{Graf00} to be consistent with the low
temperature power laws observed in specific heat and thermal
conductivity experiments. Our conclusions are, however,
insensitive  to the detailed form of the gap function for triplet
pairing. We take $\Delta_0 \approx 1$ meV as reported by Andreev
point-contact spectroscopy \cite{Lau00}.

For spin triplet pairing, and isotropic spin fluctuations,
the unrenormalized band susceptibility,
$\chi$, is given by \cite{Bri74}
\begin{eqnarray}
\chi^{rs}_{ij}({\bf p})&=& -{1 \over 2} \sigma^{i}_{\zeta\eta}
\sigma^{j}_{\tau\delta} T \sum_{{\bf k},m} \ {\cal A}^{rs}_{\bf
k,q} \ \Big\{ G^r_{\eta\tau}({\bf l}) G^s_{\delta\zeta}({\bf l+p})
\nonumber \\ & & \qquad\qquad
- [F^{r}_{\zeta\tau}({\bf l})]^*
F^s_{\eta\delta}({\bf l+p}) \Big\} \label{chi0}
\, ,
\end{eqnarray}
where $r,s=\alpha,\beta,\gamma$ are band indices, ${\bf p}=({\bf
q},i \omega_n)$, ${\bf l}=({\bf k},i \nu_m)$ are four-vectors, and
\begin{equation}
G^r_{\eta \tau}({\bf l}) = - \delta_{\eta \tau} { i \nu_m +
\epsilon_r({\bf k}) \over \nu_m^2 + E^2_r({\bf k}) } \,,
\hspace{0.2cm} F^r_{\eta \tau} ({\bf l}) = {
\Delta_{\eta \tau}({\bf k}) \over \nu_m^2 + E^2_r({\bf k}) }
\,,
\label{GF}
\end{equation}
are the normal and anomalous Greens functions, respectively, with
$E_r({\bf k})=\sqrt{\epsilon^2_r({\bf k})+ |\Delta_{\bf k}|^2}$ \cite{com6}.
The hybridization between the bands is reflected in
\begin{eqnarray}
{\cal A}^{rs}_{\bf k,q}&=& {1 \over 2} \pm { \epsilon^{-}_{\bf k} \,
\epsilon^{-}_{\bf k+q} + t_\perp^2 \over 2\sqrt{(\epsilon^{-}_{\bf
k})^2+t_\perp^2} \sqrt{(\epsilon^{-}_{\bf k+q})^2+t_\perp^2} } \,
,
\end{eqnarray}
where the upper (lower) sign applies to
$rs=\alpha\alpha,\beta\beta$ ($rs=\alpha\beta,\beta\alpha$),
${\cal A}^{\gamma\gamma}=1$, and ${\cal A}^{rs}=0$ otherwise.
In what follows we distinguish between
$\chi^{hyb}_{ij}=\chi^{\alpha\alpha}_{ij}+\chi^{\beta\beta}_{ij}
+2\chi^{\alpha\beta}_{ij}$, which arises from intra- and interband
quasiparticle transitions in the $\alpha$ and $\beta$-bands, and
$\chi^{\gamma}_{ij}\equiv \chi^{\gamma\gamma}_{ij}$ due to
quasiparticle excitations in the $\gamma$-band. Note that the
out-of-plane, $\chi_{zz}({\bf p})$, and in-plane susceptibility,
$\chi_{\pm}({\bf p})$, differ in the form of their superconducting
coherence factors, which as we show below, gives rise to their
{\it qualitatively} different frequency and momentum dependence.
Finally, the bare susceptibility, Eq.(\ref{chi0}), in correlated
electron systems is renormalized by an effective quasiparticle
interaction, $U$, and one has in random-phase approximation (RPA),
neglecting vertex corrections
\begin{equation}
{\overline \chi}^{hyb,\gamma}_{ij}=  \chi^{hyb,\gamma}_{ij}
\left( 1-U \chi^{hyb,\gamma}_{ij} \right)^{-1} \ .
\label{RPA}
\end{equation}

%
%

\begin{figure} [t]
\begin{center}
\leavevmode
\epsfxsize=7.0cm
\epsffile{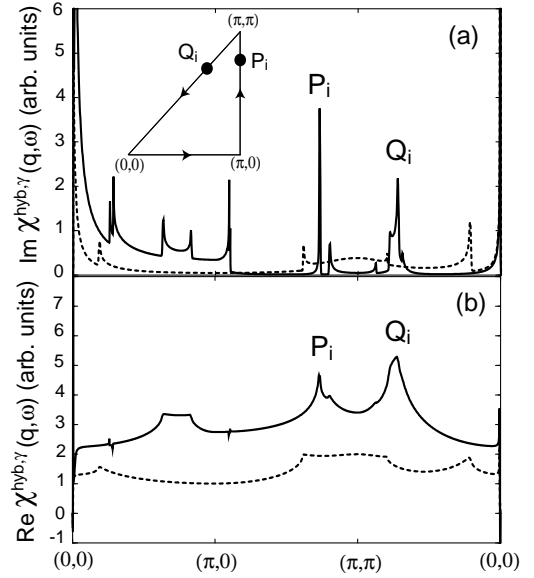}
\end{center}
\caption{ {\bf q}-scans of {(a)} Im$\, \chi^{i}_{NS}$, and  { (b)}
Re$\, \chi^{i}_{NS}$  for $i=hyb$ (solid line) and
$i=\gamma$ (dashed line) at  $\omega=6.0$ meV and $T=1.0$ meV.
Inset (a): Path of {\bf q}-scan with
filled circles showing wavevectors ${\bf Q}_i$ and ${\bf P}_i$. }
\label{chi0q}
\end{figure}

In Fig.~\ref{chi0q} we present the normal state susceptibility,
$\chi_{NS}= (\chi_{zz}+2\chi_{\pm})/3$, obtained from
Eq.(\ref{chi0}) with $\Delta_0=0$ for $\omega=6.0$ meV along the
momentum path shown in the inset.
In the vicinity of $(\pi,\pi)$, $\chi^{hyb}_{NS}$ exhibits
peaks at ${\bf Q}_i$ and ${\bf P}_i$, arising from the nesting properties of the
$\alpha$ and $\beta$-bands, while $\chi^{\gamma}_{NS}$
provides only a weakly ${\bf q}$-dependent background \cite{com3}.
Moreover, for $q \rightarrow 0$ the form of Im$\, \chi^{hyb}_{NS} \sim q^{-1}$
reflects the predominantly one-dimensional (1D) character of the
$xz,yz$-bands, while Im$\, \chi^{\gamma}_{NS} \sim \omega/q$
arises from the cylindrical nature of the $xy$-band.

\begin{figure} [t]
\begin{center}
\leavevmode
\epsfxsize=7.0cm
\epsffile{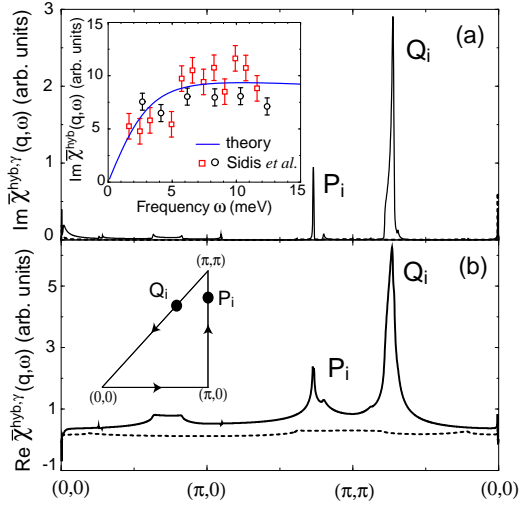}
\end{center}
\caption[]{ {\bf q}-scans for ${\overline \chi}^{hyb}_{NS}$ (solid
line) and ${\overline \chi}^{\gamma}_{NS}$ (dashed
line) for the same parameters
as shown in Fig.~\ref{chi0q}. Inset (a): Fit of Im$\, {\overline
\chi}^{hyb}$ at ${\bf Q}_i$ to the data of
Ref.~\cite{Sid99}; Im$\, {\overline \chi}$ is multiplied by a mass
enhancement factor $m^*/m_{\rm band} \sim 4$ in agreement with
dHvA experiments \cite{Mac96,Ber00}.}
\label{chi_rpa}
\end{figure}

In Fig.~\ref{chi_rpa} we present the RPA susceptibility,
${\overline \chi}_{NS}$, in the normal state. A fit of our
results, Eq.(\ref{RPA}), to the measured $\omega$-dependence of
Im$\, \chi_{NS}$ at ${\bf Q}_i$ (see inset) yields $U=0.175$ eV
\cite{com5} in agreement with Ref.~\cite{Maz99}. Due to the {\bf
q}-structure of Re$\, \chi^{hyb}_{NS}$ (Fig.~\ref{chi0q}b), and
the weak {\bf q}-dependence of $U$ \cite{Maz99}, Im$\, {\overline
\chi}^{hyb}_{NS}$ is reduced from its bare value for small {\bf
q}, but still possesses peaks at ${\bf Q}_i$ and ${\bf P}_i$. In
contrast, $\, {\overline \chi}^{\gamma}_{NS}$ is strongly
suppressed for all {\bf q}. Thus, the experimentally observed peak
close to ${\bf Q}_i$ arises primarily from Im$\, {\overline
\chi}^{hyb}_{NS}$ and the strongest SC pairing most likely occurs
between electrons in the $\beta$-band.

%
%

\begin{figure} [t]
\begin{center}
\leavevmode
\epsfxsize=7.0cm
\epsffile{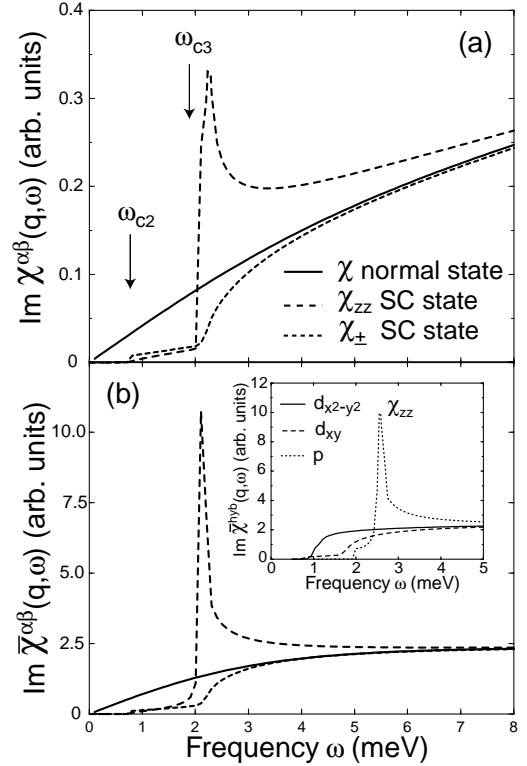}
\end{center}
\caption{Spin susceptibilities at ${\bf Q}_i$ for the $f_{xy}$-wave
state at $T=0$:
{(a)} bare susceptibility, Im$\, \chi^{hyb}$; {(b)}
RPA susceptibility, Im$\, {\overline \chi}^{hyb}$,
for $U=0.175\,$eV.
The frequency integral of Im$\, \chi^{hyb}({\bf Q}_i)$
up to 15\ meV remains constant through $T_c$.
Inset: For spin singlet states with
$d_{xy}$ or $d_{x^2-y^2}$ symmetry Im$\, {\overline \chi}^{hyb}_{zz}$ shows
{\it no} resonance peak, contrary to the spin triplet $p$-wave state.}
\label{chiom}
\end{figure}

In Fig.~\ref{chiom}a we present the frequency dependence of Im$\,
\chi^{hyb}$ at ${\bf Q}_i$ in the normal and superconducting
state. There exist three channels for quasiparticle excitations
with wavevector ${\bf Q}_i$ which contribute to Im$\, \chi^{hyb}$,
as indicated by arrows in Fig.~\ref{FS}. In the normal state all
three channels are excited in the low frequency limit, which
yields Im$\, \chi^{hyb}_{NS} \sim \omega$, in agreement with our
numerical results in Fig.~\ref{chiom}a. The dominant contribution
to Im$\, \chi^{hyb}$, both in the normal and superconducting
state, arises from excitations of type (3), since (a) they are
intraband $xz$ (or $yz$) transitions and thus independent of
$t_\perp$, and (b) the FS exhibits the largest nesting in this
region of momentum space. 

In the superconducting state excitations (1-3) possess nonzero
threshold energies, $\omega_{cn}$ with $n=1,2,3$,
that are determined by the
momentum dependence of the order parameter and the shape of the
Fermi surface.
Specifically, $\omega_{cn}=|\Delta_{\bf k}|+|\Delta_{\bf k+Q_i}|$,
where ${\bf k}$ and ${\bf k+Q}_i$ both
lie on the Fermi surface, as shown in Fig.~\ref{FS}. For the band
parameters chosen, we obtain $\omega_{c1} \simeq 0.15
\Delta_0$, $\omega_{c2} \simeq 0.8 \Delta_0$, and
$\omega_{c3} \simeq 2.1 \Delta_0$. Since excitations (1-3) are
well separated in frequency, we can identify their relative
contribution to Im$\, \chi^{hyb}_{zz,\pm}$. While $\omega_{c1}$
cannot be observed in the frequency dependence of Im$\,
\chi^{hyb}_{zz,\pm}$ due to the negligible spectral weight of
excitation (1), $\omega_{c2}$ and $\omega_{c3}$
can clearly be identified. The large spectral weight of
excitation (3) likely makes $\omega_{c3}$ the experimentally
observable spin gap. Moreover, due to the superconducting
coherence factors which appear in the calculation of
$\chi^{hyb}_{zz, \pm }$, the overall frequency dependence of the
in-plane and out-of-plane component of Im$\, \chi^{hyb}$ are {\it
qualitatively} different. Specifically, since Re$(\Delta_{\bf k}
\, \Delta^*_{\bf k+q})$ is negative for transition (3), but
positive for transition (2), Im$\, \chi^{hyb}_{zz}$ (Im$\,
\chi^{hyb}_{\pm}$ ) exhibits a sharp jump at
$\omega_{c3}$ ($\omega_{c2}$) and increases
continuously at $\omega_{c2}$ ($\omega_{c3}$).
Consequently, Re$\, \chi^{hyb}_{zz}$ (Re$\, \chi^{hyb}_{\pm}$)
possesses a logarithmic divergence at
$\omega_{c3}$ ($\omega_{c2}$).

%
%

In Fig.~\ref{chiom}b we present the RPA susceptibility, Im$\,
{\overline \chi}^{hyb}_{zz,\pm}$, in the superconducting state,
assuming that $U$ remains unchanged below $T_c$. Due to the
logarithmic divergence of Re$\, \chi^{hyb}_{zz}$ at $\omega_{c3}$,
Im$\, {\overline \chi}^{hyb}_{zz}$ exhibits a {\it resonance} peak
at a frequency slightly below $\omega_{c3}$. In contrast, Im$\,
{\overline \chi}^{hyb}_{\pm}$ increases continuously above
$\omega_{c3}$. The logarithmic divergence of Re$\,
\chi^{hyb}_{\pm}$ at $\omega_{c2}$ is rapidly smoothed out for
finite quasiparticle damping due to its small prefactor and is
likely experimentally not observable. Thus, we predict that for
triplet pairing Im$\, {\overline \chi}^{hyb}_{zz}$ and Im$\,
{\overline \chi}^{hyb}_{\pm}$ possess {\it qualitatively}
different frequency dependencies at ${\bf Q}_i$ with only Im$\,
{\overline \chi}^{hyb}_{zz}$ exhibiting a resonance peak below
$\omega_{c3}$. In contrast, a resonance peak was predicted  in
Refs.~\cite{Kee00,Fay00} for the in-plane component Im$\,
{\overline \chi}_{\pm}$, but not for Im$\, {\overline \chi}_{zz}$.
A comparison of our results for $\chi_{zz,\pm}$ with those in
\cite{Kee00,Fay00} suggests that the SC coherence factors for
$\chi_{zz,\pm}$ have been interchanged in
Refs.~\cite{Kee00,Fay00}. 
We obtain the correct $\omega, q \rightarrow 0$ limit
only for the SC coherence factors which
appear in our results for $\chi_{zz,\pm}$ in  Eq.(\ref{chi0}).
In this case,
we find that Re$\, \chi_{zz}$ decreases below $T_c$
when a SC gap opens, while Re$\, \chi_{\pm}$ remains unchanged. As
shown by Leggett \cite{Leg75}, this result is a  general property
of any unitary state if ${\bf d} || \hat{c}$.

We find that our results are insensitive to details of the
electronic band structure or the symmetry of the gap function for
spin triplet pairing. In particular, for a nodeless
superconducting gap with `$p$-wave' symmetry \cite{Rice95},
$\Delta({\bf k})=\Delta_0 \left(\sin{k_x}+i \sin{k_y}\right)$,
belonging to the $E_u$ representation, the frequency and momentum
dependence of Im$\, {\overline \chi}^{hyb}_{zz,\pm}$  remains to a
large extent unchanged from that shown in Fig.~\ref{chiom}b (see
inset); a resonance peak appears again only in Im$\,
{\overline \chi}^{hyb}_{zz}$. In contrast, for spin singlet
pairing the in-plane and out-of-plane susceptibilities are
identical and our calculations (analogous to triplet pairing) are
{\it qualitatively} different since no resonance peak exists in
Im$\, {\overline \chi}^{hyb,\gamma}$. In the inset of
Fig.~\ref{chiom} we plot Im$\, {\overline \chi}^{hyb}$ at ${\bf
Q}_i$ as a function of frequency for SC gaps with $d_{x^2-y^2}$
symmetry, $ \Delta({\bf k})=\Delta_0 \left(\cos{k_x}-
\cos{k_y}\right)/2$, and $d_{xy}$-symmetry, $\Delta({\bf
k})=\Delta_0 \sin{k_x}\sin{k_y}$, with $\Delta_0=1$ meV. In both
cases, Im$\, {\overline \chi}^{hyb}$ increases continuously above
$\omega_{c3}$, since $\Delta({\bf k})$ does {\it not} change sign
for excitation (3) and no logarithmic singularity occurs in Re$\,
\chi^{hyb}$. In contrast, for the FS geometry of the HTSC and a SC
gap with $d_{x^2-y^2}$ symmetry, one finds $\Delta_{\bf k} \,
\Delta_{\bf k+Q}<0$, which as described above leads to a resonance
peak at ${\bf Q}=(\pi,\pi)$ \cite{Maz95}. A resonance peak is thus
{\em not} an intrinsic property of singlet or triplet
superconductivity, but arises from the interplay of FS topology
and symmetry of the SC gap.

An additional contribution to $\chi_{\pm}$ in the SC state can in
principle come from a coupling of the spin density to in-plane
fluctuations of {\bf d}. However, for the {\bf q}-independent
coupling assumed in Ref.~\cite{Kee00}, we find that these
fluctuation contributions (FC) are three orders of magnitude
smaller than those coming from Eq.(\ref{chi0}). Moreover, the
spin-orbit coupling present in Sr$_2$RuO$_4$ introduces a gap for
in-plane fluctuations of {\bf d} which further suppresses the FC
to $\chi_{\pm}$ and renders them irrelevant.

In summary, we present a scenario for the spin susceptibility in
the normal and SC state of Sr$_2$RuO$_4$. In the normal state we
find a peak close to the experimentally observed position at ${\bf
Q}_i$. For spin triplet pairing in the superconducting state we
show that the momentum and frequency dependence of Im$\,
{\overline \chi}_{zz}$ and Im$\, {\overline \chi}_{\pm}$ are {\it
qualitatively} different. We predict the appearance of a resonance
peak in Im$\, {\overline \chi}_{zz}$, similar to the one observed
in the HTSC, and its absence in Im$\, {\overline \chi}_{\pm}$.
Finally, we show that no resonance peak exists for spin singlet
pairing.

We would like to thank A.V. Balatsky, P. Dai and Y. Maeno for stimulating
discussions. This work was supported in part through the Los
Alamos Summer School and the Department of Energy.


\begin{thebibliography}{99}
\bibitem{com2}
The coupling of the $\gamma$-band to the $\alpha,
\beta$-bands, implied by observation of a single $T_c$, is at the
most weak and irrelevant for our results.
\bibitem{com1}
We neglect the weaker nearest neighbor hopping terms between
the $xz$ and $yz$-bands.
\bibitem{com6}
We neglect interband pairing since it leads to non-zero total
momentum  of the Cooper pairs.
\end{thebibliography}
\end{document}